\documentclass{ws-procs975x65}

\usepackage{cancel}

\newcommand{\be}{\ensuremath{\beta} }
\newcommand{\la}{\ensuremath{\lambda} }

\newcommand{\lsim}{\ensuremath{\lesssim} }
\newcommand{\X}{\ensuremath{\!\times\!} }
\newcommand{\MSbar}{\ensuremath{\overline{\textrm{MS}} } }
\newcommand{\fig}[1]{Fig.~\ref{#1}}

\newcommand{\Sb}{\ensuremath{\cancel{S^4}} }

\begin{document}
\title{Finite-temperature study of eight-flavor SU(3) gauge theory}
\author{David~Schaich,$^{1*}$ Anna Hasenfratz$^2$ and Enrico Rinaldi$^3$ \\ for the Lattice Strong Dynamics (LSD) Collaboration}
\address{$^1$Department of Physics, Syracuse University, Syracuse, New York 13244, United States \\
         $^2$Department of Physics, University of Colorado, Boulder, Colorado 80309, United States \\
         $^3$Lawrence Livermore National Laboratory, Livermore, California 94550, United States \\
         $^*$E-mail: dschaich@syr.edu}

\begin{abstract} 
  We present new lattice investigations of finite-temperature transitions for SU(3) gauge theory with $N_f = 8$ light flavors.
  Using nHYP-smeared staggered fermions we are able to explore renormalized couplings $g^2 \lsim 20$ on lattice volumes as large as $48^3\X 24$.
  Finite-temperature transitions at non-zero fermion mass do not persist in the chiral limit, instead running into a strongly coupled lattice phase as the mass decreases.
  That is, finite-temperature studies with this lattice action require even larger $N_T > 24$ to directly confirm spontaneous chiral symmetry breaking. \\
\end{abstract}

\bodymatter 
\noindent SU(3) gauge theory with $N_f = 8$ massless fundamental flavors is currently the subject of considerable interest, both as a quantum field theory exhibiting strongly coupled dynamics significantly different from QCD, and also as the basis for models of new strong dynamics producing a standard-model-like 125-GeV Higgs particle.
We can only include here an incomplete collection of references to recent investigations employing both continuum and lattice methods.\cite{Appelquist:2009ty, Schaich:2012fr, Bashir:2013zha, Hasenfratz:2013uha, Schaich:2013eba, Aoki:2014oha, Kurachi:2014qma, Appelquist:2014zsa, Lombardo:2014mda, Hasenfratz:2014rna, Fodor:2015baa}
In this work we attempt to confirm the conventional wisdom that chiral symmetry breaks spontaneously for $N_f = 8$, which would rule out the existence of an 8-flavor conformal IR fixed point (IRFP).
However, this requires extrapolating to the massless chiral limit, and we are unable to establish that chiral symmetry breaking persists in that limit.

Previous lattice studies have explored the discrete \be function of the 8-flavor system, denoted $\be_s(g^2)$ for scale change $s$.\cite{Appelquist:2009ty, Hasenfratz:2014rna, Fodor:2015baa}
An IRFP would correspond to $\be_s(g_{\star}^2) = 0$, and no such zero has been observed.
In fact, the \be function was found to be monotonic throughout the ranges of couplings explored: $g_{SF}^2 \lsim 6.6$ in the Schr\"odinger functional scheme,\cite{Appelquist:2009ty} $g_c^2 \lsim 6.3$ in a gradient flow scheme with $c = 0.3$,\cite{Fodor:2012td, Fodor:2015baa} and $g_c^2 \lsim 14$ in gradient flow schemes with $c = 0.25$ and $c = 0.3$.\cite{Hasenfratz:2014rna}
The last result is the most relevant to this work, since we use the same lattice action with nHYP-smeared staggered fermions and both fundamental and adjoint plaquette terms with couplings related by $\be_A / \be_F = -0.25$.\cite{Cheng:2011ic}
This enables us to relate our bare lattice couplings $\be_F$ to renormalized $g_c^2$.
This work is part of ongoing investigations by the Lattice Strong Dynamics (LSD) Collaboration, using this action to extend the USBSM project.\cite{Schaich:2013eba}

The non-observation of an IRFP by lattice studies of the discrete \be function does not guarantee that the 8-flavor theory exhibits spontaneous chiral symmetry breaking.
On their own such calculations cannot rule out the possibility that the system flows to an IRFP at a stronger coupling.
In fact, the non-perturbative $\be_s(g_c^2 \approx 14)$ is comparable to the four-loop \MSbar prediction,\cite{Hasenfratz:2014rna} which does possess an IRFP at $g_{\MSbar}^2 \approx 19.5$. 
To exclude such behavior one must demonstrate that the massless system spontaneously breaks chiral symmetry at some $g^2$ for which the \be function is still non-zero.
Here we attempt to do this by studying chiral symmetry breaking at finite temperature $T = 1 / (aN_T)$ and non-zero fermion mass $am$, where ``$a$'' is the lattice spacing and the lattice volume is $L^3\X N_T$ with $L / N_T = 2$.\cite{Schaich:2012fr, Hasenfratz:2013uha}
We work with fixed $N_T = 20$ and 24 for small $0.0025 \leq am \leq 0.01$, and extrapolate $am \to 0$ to investigate the massless chiral limit.

\begin{figure}[bp]
  \centering
  \includegraphics[width=0.45\linewidth]{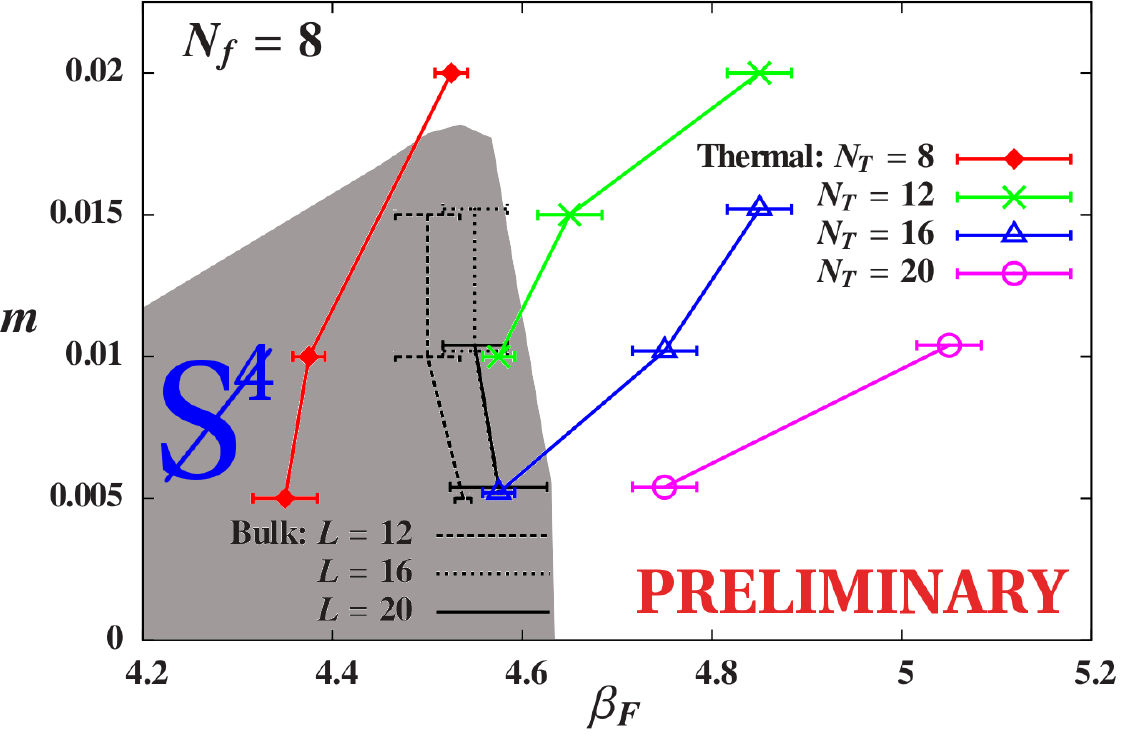}\hfill
  \includegraphics[width=0.45\linewidth]{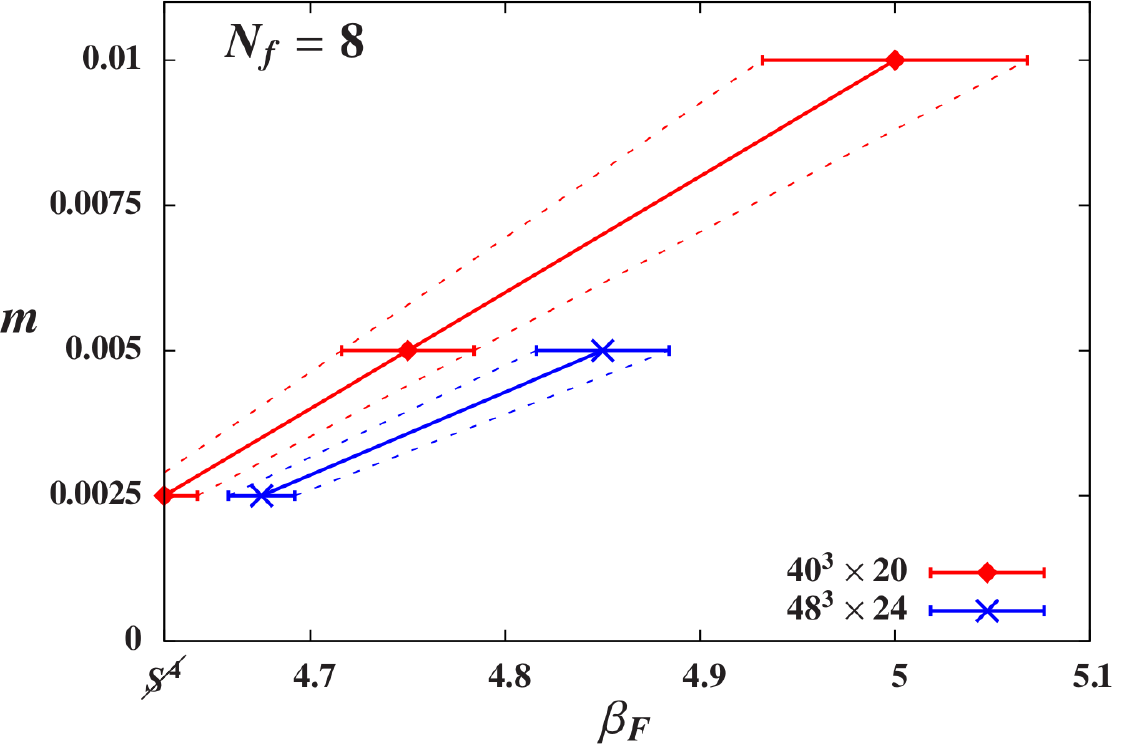}
  \caption{\label{fig:phase}{\bf Left:} Previous $N_f = 8$ studies found that the finite-temperature transitions merge with zero-temperature bulk transitions into the \Sb lattice phase as the fermion mass $am$ decreases.\protect\cite{Schaich:2012fr, Hasenfratz:2013uha}  {\bf Right:} Our new results produce the same behavior at smaller masses on larger lattice volumes, implying that $N_T > 24$ is required to establish spontaneous chiral symmetry breaking for $N_f = 8$.  This plot zooms in on the weak-coupling regime to the right of the \Sb phase.}
\end{figure}

The chiral extrapolation is crucial since $am > 0$ explicitly breaks chiral symmetry and can even produce QCD-like scaling that disappears as $am$ decreases.
This behavior was observed in previous studies (using the same lattice action) that considered $am \geq 0.005$ and $N_T \leq 20$.\cite{Schaich:2012fr, Hasenfratz:2013uha}
As shown in the left plot of \fig{fig:phase}, QCD-like scaling between $12 \leq N_T \leq 16$ for $am \geq 0.01$ is lost at $am = 0.005$ where the finite-temperature transitions merge with bulk (zero-temperature) transitions into the ``$\Sb$'' lattice phase in which the single-site shift symmetry ($S^4$) of the staggered action is spontaneously broken.\cite{Cheng:2011ic}
In IR-conformal systems such behavior persists in the chiral limit as $N_T \to \infty$, whereas for chirally broken systems the massless transitions must move to $\be_F^{(c)} \to \infty$ as $N_T \to \infty$.

The right plot of \fig{fig:phase} shows our new results, which include smaller $am = 0.0025$ on $40^3\X 20$ and $48^3\X 24$ lattices.
These $N_T = 20$ and 24 are extraordinarily large compared to typical lattice QCD calculations.
Unfortunately they do not suffice to establish spontaneous chiral symmetry breaking.
At $am = 0.0025$ the $N_T = 20$ finite-temperature transition also merges with the bulk transition into the \Sb lattice phase.
Even the $N_T = 24$ finite-temperature transitions will clearly run into the \Sb phase at non-zero mass, rather than reaching the chiral limit.

\begin{figure}[bp]
  \includegraphics[width=0.45\linewidth]{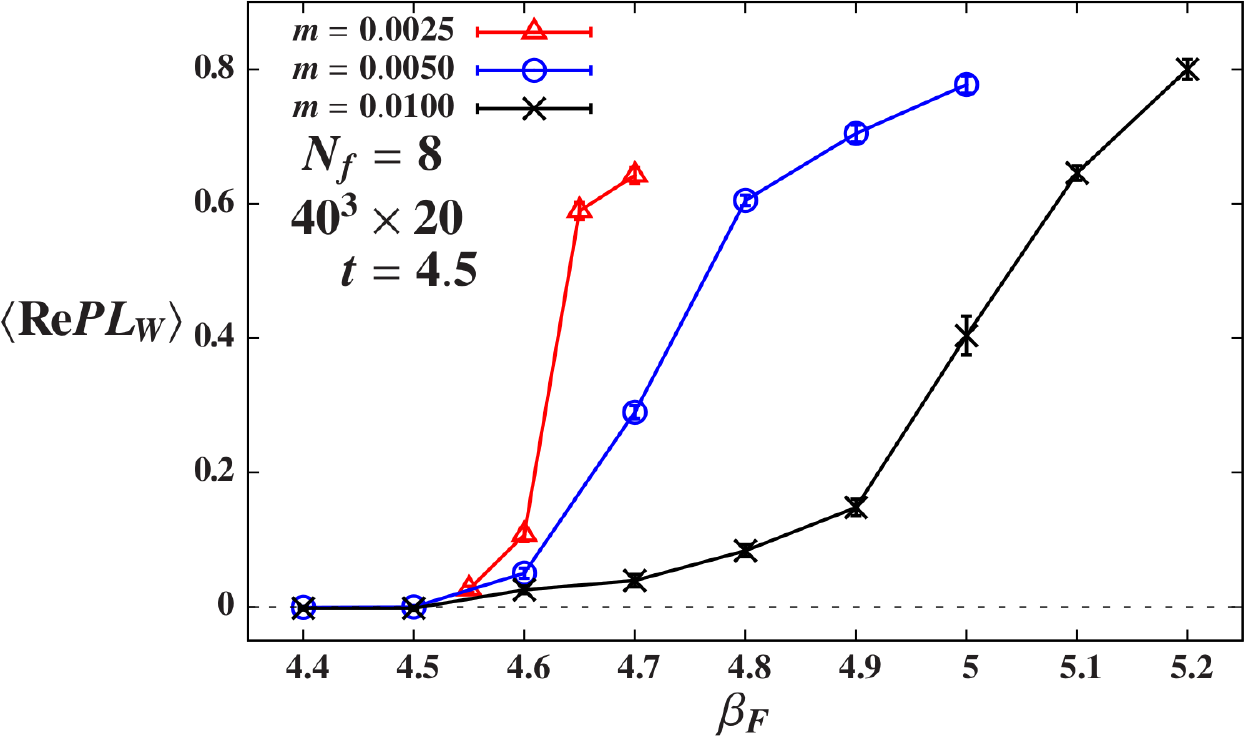}\hfill
  \includegraphics[width=0.45\linewidth]{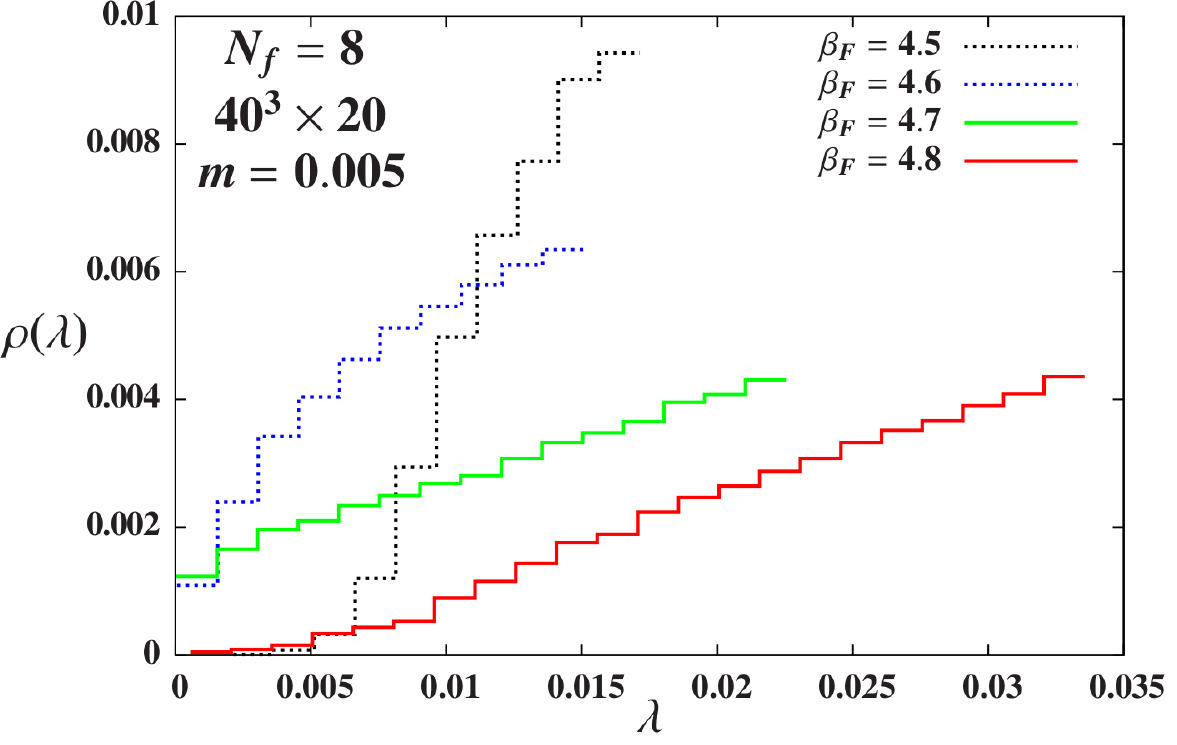}
  \caption{\label{fig:Wpoly}{\bf Left:} The $c = 0.3$ Wilson-flowed Polyakov loop $PL_W$ for $40^3\X 20$ lattices with $am = 0.01$, 0.005 and 0.0025 vs.\ the bare lattice coupling $\be_F$.  As the fermion mass decreases the transitions sharpen and move to stronger coupling, eventually merging with the zero-temperature bulk transition into the \Sb phase at $\be_F \approx 4.625$.  {\bf Right:} The massless Dirac eigenvalue spectrum $\rho(\la)$ contrasts the three phases encountered for $40^3\X 20$ lattices with $am = 0.005$: the chirally symmetric phase at weak coupling ($\be_F = 4.8$), the chirally broken phase at intermediate coupling ($\be_F = 4.7$) and the \Sb lattice phase at strong coupling ($\be_F = 4.5$).  We directly measure 200 eigenmodes to produce each histogram.}
\end{figure}

In \fig{fig:Wpoly} we present some results for observables that have proven useful to identify both bulk and finite-temperature transitions: the Wilson-flowed Polyakov loop $PL_W$ and the massless Dirac eigenvalue spectrum $\rho(\la)$.
$PL_W$ is a modern adaptation of the RG-blocked Polyakov loop investigated in previous studies, which significantly improves signals of the finite-temperature transition without altering its location.\cite{Schaich:2012fr, Hasenfratz:2013uha} 
It is trivial to measure the Polyakov loop as a function of Wilson flow time $t$, and sufficiently large $t$ produces a clear contrast between confined systems with small $PL_W$ and deconfined systems with large $PL_W$.
This is shown in the left plot of \fig{fig:Wpoly} for $N_T = 20$ and $t = (0.3N_T)^2 / 8$ corresponding to $c = \sqrt{8t} / N_T = 0.3$.
As the fermion mass decreases from $am = 0.01$ the finite-temperature transition in $PL_W$ steadily sharpens and moves to stronger coupling, merging with the bulk transition into the \Sb phase at $am = 0.0025$ as in \fig{fig:phase}.

The right plot of \fig{fig:Wpoly} shows the eigenvalue spectrum $\rho(\la)$ for a subset of the $N_T = 20$ ensembles with $am = 0.005$, clearly contrasting the three different phases we can observe for this mass.
At the weakest coupling shown, $\be_F = 4.8$, the system is deconfined and chirally symmetric, with $\rho(0) = 0$ and a gap below the smallest eigenvalue $\la_0 > 0$.
The gap grows at even weaker couplings that are not included in this plot.
Moving to stronger couplings, at $\be_F = 4.7$ we observe the expected chiral symmetry breaking, with $\rho(0) \neq 0$ and a small slope $\frac{d\rho}{d\la}$.
That is, we find $4.7 < \be_F^{(c)} < 4.8$ for the $N_T = 20$ transition with $am = 0.005$, slightly sharper than the signal in $PL_W$.
However, at the strongest coupling shown, $\be_F = 4.5$, chiral symmetry breaking is lost ($\la_0 > 0$) and the system exhibits the ``soft edge'' $\rho(\la) \propto \sqrt{\la - \la_0}$ characteristic of the \Sb phase.\cite{Cheng:2011ic}
Finally, $\be_F = 4.6$ appears to exhibit partial features of both the chirally broken and \Sb phases, with $\rho(0) \neq 0$ but a much larger slope approaching the square-root behavior of the soft edge.

While this finite-temperature study with $N_T = 20$ and 24 is not able to establish spontaneous chiral symmetry breaking for $N_f = 8$, it is only part of ongoing investigations by the LSD Collaboration that primarily focus on the zero-temperature hadron spectrum, and in particular the scalar Higgs particle.
When that work is finalized, combining these complementary studies of the discrete \be function, finite-temperature transitions, and hadron spectrum, all using the same lattice action, will shed further light on $N_f = 8$ and its phenomenological viability as the basis for new strong dynamics beyond the standard model. \\[-6 pt]

\noindent {\em Acknowledgments:} We thank the other members of LSD for continuing collaboration on $N_f = 8$.
AH and DS are grateful for the hospitality of the Aspen Center for Physics, supported by the U.S.~National Science Foundation under Grant No.~PHYS-1066293.
This work was supported by Lawrence Livermore National Laboratory (LLNL) LDRD~13-ERD-023 (ER) and by the U.S.~Department of Energy (DOE), Office of Science, Office of High Energy Physics, under Award Numbers {DE-SC}0010005 (AH), {DE-SC}0008669 (DS) and {DE-SC}0009998 (DS). 
Numerical calculations were carried out through LLNL Institutional Computing Grand Challenge program allocations on the LLNL BlueGene/Q (rzuseq and vulcan) supercomputer and on the DOE-funded USQCD facilities at Fermilab.

\raggedright
\bibliographystyle{ws-procs975x65}
\bibliography{SCGT15}
\end{document}